# Doping induced Mott collapse and the density wave instabilities in $(Sr_{1-x}La_x)_3Ir_2O_7$

Zhenyu Wang[1][†], Daniel Walkup[2,3][†], Yulia Maximenko[1], Wenwen Zhou[2], Tom Hogan[2,4], Ziqiang Wang[2], Stephen D. Wilson[4] and Vidya Madhavan[1]*

[1]Department of Physics and Frederick Seitz Materials Research Laboratory, University of Illinois Urbana-Champaign, Urbana, Illinois 61801, USA

[2]Department of Physics, Boston College, Chestnut Hill, Massachusetts 02467, USA

[3]Center for Nanoscale Science and Technology, National Institute of Standards and Technology, Gaithersburg, MD 208993

[4]Materials Department, University of California, Santa Barbara, California 93106, USA

[†]equal contribution
* vm1@illinois.edu

**The path from a Mott insulating phase to high temperature superconductivity encounters a rich set of unconventional phenomena involving the insulator-to-metal transition (IMT) such as emergent electronic orders and pseudogaps that ultimately affect the condensation of Cooper pairs. A huge hindrance to understanding the origin of these phenomena in the curates is the difficulty in accessing doping levels near the parent state. Recently, the $J_{eff}$=1/2 Mott state of the perovskite strontium iridates has revealed intriguing parallels to the cuprates, with the advantage that it provides unique access to the Mott transition. Here, we exploit this accessibility to study the IMT and the possible nearby electronic orders in the electron-doped bilayer iridate $(Sr_{1-x}La_x)_3Ir_2O_7$. Using spectroscopic imaging scanning tunneling microscopy, we image the La dopants in the top as well as the interlayer SrO planes. Surprisingly, we find a disproportionate distribution of La in these layers with the interlayer La being primarily responsible for the IMT, thereby revealing the distinct site-dependent effects of dopants on the electronic properties of bilayer systems. Furthermore, we discover the coexistence of two electronic orders generated by electron doping: a unidirectional electronic order with a concomitant structural distortion; and local resonant states forming a checkerboard-like pattern trapped by La. This provides evidence that multiple charge orders may exist simultaneously in Mott systems, even with only one band crossing the Fermi energy.**

**Introduction**

When holes or electrons are introduced into a Mott insulator, such as the parent compound of high temperature superconducting cuprates, various competing electronic orders emerge and lead to a remarkably rich phase diagram (1). Despite years of intense studies, there is still no consensus on the origin of these orders and the relationship between them. This is partly because of the difficulties in obtaining high quality crystals in the lower doping regime close to the Mott transition, as well as the highly insulating nature of Mott systems, which makes them prohibitive for local probes like scanning tunneling microscopy (STM). Recently, an intriguing parallel has arisen between the S=1/2 Mott phases of the cuprate and the spin-orbit coupling driven $J_{eff}$=1/2 Mott states of their heavier, iridium oxide cousins (2-14), which opens up a new pathway to unraveling some of the mysteries of the lightly doped Mott insulators.

The Ruddlesden-Popper (RP) series of strontium iridates is a homologous series of compounds containing corner sharing $IrO_6$ octahedra formed into planes and separated by rock-salt SrO spacing layers. The repeating number of layers of $IrO_6$ octahedra increases from one to infinity as the series progresses from the single layer, two-dimensional limit, to the fully three-dimensional perovskite structure in the infinite layer limit. Compared to the single layer $Sr_2IrO_4$ (Ir214), the next member of the series, $Sr_3Ir_2O_7$ (Ir327), has a more complicated crystal structure: it is composed of stacked bilayer-$IrO_6$-octahedra, which share one SrO charge reservoir layer in between. The multiplicity of charge reservoir layers offers a new variation in the dopant position whose effects have not yet been studied at the microscopic scale. Specifically, an A-site dopant on the top layer is in a very different environment from an A-site dopant in the second layer (Sr1 and Sr2 in Fig.1a).

Ir327 retains the $J_{eff}$=1/2 antiferromagnetic Mott state with a smaller change gap (14, 15) of 130meV, making it suitable for STM studies. When $Sr^{2+}$ is substituted with $La^{3+}$, the system undergoes a first order transition to a correlated metallic state (16-18), where a structural distortion and a low energy excitation suggesting the existence of an electronic density wave (DW) have been recently detected (14, 19). Yet, little is known about the emergence of this novel metallic state from the weak Mott background, leaving many open questions, such as: How do the $La^{3+}$ dopants change the electronic structure at the atomic scale? How do the electrons arrange themselves in the phase-separated state? What symmetries are broken in the DW state? In this work, we use scanning tunneling microscopy and spectroscopy to study $(Sr_{1-x}La_x)_3Ir_2O_7$ with x= 0.032, 0.039 and 0.048 near the insulator-metal transition (IMT) and visualize two kinds of electronic DWs. In addition, we find that the two inequivalent Sr sites have surprisingly distinct effects on the electronic properties: it is predominantly the interlayer La that induce the IMT, which proceeds through a transfer of local spectral weight into the Mott gap.

**Results**

The crystals cleave easily between two adjacent SrO layers, exposing SrO terminated surfaces. Figure 1b depicts a topographic image taken on the x=0.032 sample, and it represents several typical features in these compounds: A square lattice is visible with lattice constant a~3.9 Å, and atomic scale features in the form of bright squares of size 2a (see Supplemental Material) are also observed, which were absent in previously studied parent samples (15). We identify these as La dopants. Note that since La substitutes for Sr, and the center of the dopants lie on the lattice sites observed in the images, we conclude that the lattice of atoms seen in the topography is the Sr lattice. These La atoms substituting for the top layer Sr1 sites are henceforth called La1. Zooming into the La dopants the density of states around each defect is in fact dimorphic, appearing like two parallel lines instead of a square (Fig. 1c and d), which reflects the symmetry of the underlying lattice. As shown in Fig. 1a, the staggered rotation of the $IrO_6$ octahedra about c-axis make two adjacent Sr sites in the same SrO plane inequivalent: for one site the closest O atoms in the $IrO_2$ plane are along a- axis, while for the other they are along b- axis. Thereby, the immediate environment of a Sr site preserves the reflection symmetry, but locally reduces C4 rotation symmetry to C2. By indexing the lattice sites according to this rotation, we find that the directions of the parallel lines of high density of states near La1, are perfectly correlated with the lattice positions in this large field of view (fig. S1). This not only provides added confirmation that the observed surface atoms belong to Sr sites, but also establishes the long-range ordered nature of the octahedral rotation in this sample.

A visual examination of the topography reveals that the density of La1 does not correspond to the nominal doping. In fact, a count of the number of La1 gives an areal density corresponding to a nominal doping of only ~1.5% (assuming a uniform distribution across the three SrO layers), which is far smaller than the actual doping level of 3.2% determined via energy dispersive spectroscopy (EDS) measurements. To find the missing La atoms, we obtain atomically resolved dI/dV(r, eV) maps at different energies, as shown in Fig.2a and b. We find a forest of bright atomic-scale features appearing in the local density of states (LDOS) maps near $E_F$ (Fig. 2b), reminiscent of the previously observed oxygen dopants in bilayer cuprates (20, 21). These bright features cannot be accounted for by La1 (blue dots in Fig. 2b) and therefore represent distinct atomic scale impurities.

To clarify the chemical nature of these bright dots, we check their positions with respect to the Sr lattice. We first apply the Lawler-Fujita drift-correction algorithm (22) to the atom-resolved topographies and dI/dV maps. This process removes the warping caused by piezo relaxation and thermal drift. A zoomed-in image of the area marked by the white square in Fig. 2b, is shown in Fig. 2c. Visual inspection shows that most of the bright dots line up with the Sr lattice. To confirm this for a statistically significant number of impurities, we obtain the coordinates of the centers of all spots in a 20nm*20nm area. These are plotted with respect to the lattice in Fig. 2d. We find that within the error represented by the resolution, the vast majority of the dots are centered close to Sr sites. The position of the bright dots at Sr sites, combined with the lack of such high intensity spots in the parent compound suggests that these arise from La dopants occupying Sr sites one-layer underneath (the Sr2 site). These La dopants at the Sr2 sites are henceforth called La2.

This hypothesis that the bright dots correspond to La dopants in Sr2 sites can be confirmed by counting their number and comparing this to the expected doping concentration from EDS. Assuming all the observed bright features in the dI/dV maps are La2, we get the total doping concentration to be $x = 2/3\rho_{La1} + 1/3\rho_{La2}$. For the 3.2% sample, the areal densities for La1 and La2 substitutions are found to be 1.5% and 7.3% respectively, so one obtains x~3.4% which is within error of the EDS results of 3.2%. A similar analysis for the 3.9% compound (fig. S2) gives us a doping level of 4.1% by counting. While there is a possibility that the La2 atoms occupy an interstitial site that coincidently lies right under the Sr sites, we believe that this is unlikely based on a comparison with the La doped single layer Ir214 where a count of the top La1 seen in the topography corresponds well to the nominal doping (fig. S2). Since the only difference between the single and bilayer compounds is that the middle layer Sr2 site does not exist in the single layer, this provides further supporting evidence for La2 being substitutional dopants at the Sr2 sites. From the data, the density of La2 (7.3%) is more than double the outer two layers put-together (3%). This is an unexpected finding and indicates that La dopants preferentially occupy the Sr sites in the middle-layer of the bilayer structure.

Having identified the two doping sites leads to a natural question: Do La1 and La2 have the same effect on the density of states? Spectra on a few different La1 and La2 impurities (Fig. 2e) are shown in Fig. 2f. Comparing the spectra far from impurities (black dashed curve) to the spectra on La1 impurities (dark green), we see that La1 has little effect on the density of states except to slightly move the leading-edge position of the conduction band. The effect of La2 (pink) however, is much more dramatic, creating a large density of states within the gap (also see line-cut profiles across La1 and La2 in fig. S4).

To better visualize overall trends in the dI/dV spectra, we present 'binned-averaged spectra' in the figure 3. To obtain this, the spectra obtained in a dI/dV map are first binned against the density of states value at -40meV which well captures the inhomogeneity in the sample (see dI/dV (r, -40meV) map in Fig. 2a). We then average all spectra in each bin resulting in the binned-averaged spectra shown in figure 3. In the x=0.032 sample (Fig. 3a), the green curves corresponding to the dark regions in dI/dV(r, eV) exhibit fully insulating line shapes reminiscent of the parent compound with a gap up to 100meV. In stark contrast, the red curves show significant non-zero density of states at the Fermi energy. By comparing the shape of the red curves with spectra obtained on individual La1 and La2 impurities shown in Fig. 2f, we can see that the red curves and La2 spectra have almost identical line shapes which suggests that the subsurface La2 are responsible for the large density of states at the Fermi energy. This is confirmed by a correlation analysis between the density of La2 impurities (fig. S5) and the density of states map, which shows a correlation coefficient of 0.75. The analysis also shows negligible correlation between La1 and the density of states distribution. Thus, we deduce that the subsurface La2 are primarily responsible for the doping induced insulator metal transition.

We now turn to the question of possible charge ordering in this system. A density wave order was recently discovered in this system (14) which emerges upon electron doping and forms concomitantly with the tilting of the oxygen octahedral (19). Recent studies on the electron doped Ir214 (6) reveal a disordered density of states modulation with 2a period, which was interpreted in terms of a local stripe order resembling the 4a modulation in underdoped cuprates (23, 24). The question then arises whether a similar order is seen in $(Sr_{1-x}La_x)_3Ir_2O_7$. Examining the Fourier transforms (FTs) of the topography as well as the $dI/dV$ maps, we find that in addition to the Bragg peaks, there are peaks at the ($\pm 1/2$, $\pm 1/2$) positions, as well as a diffuse intensity around the ($\pm 1/2$, 0) and (0, $\pm 1/2$) positions (Fig. 4). The peaks at ($\pm 1/2$, $\pm 1/2$) also exist in the parent compound due to the alternating octahedral rotations and are not unique to the La doped samples. The diffuse ($\pm 1/2$, 0) signal however indicates a real space pattern with 2a periodicity along the Ir-O-Ir direction.

A visual examination of the topography suggests that the ($\pm 1/2$, 0) feature in the FT of the topography might arise from the LDOS signature of the La1 induced impurity resonance (Fig. 4a) which can be seen as a square shaped density of states of period 2a around the La1 atoms. This can be confirmed by masking these features (Fig. 4b), whereupon the ($\pm 1/2$, 0) feature in the FT of the topography disappears (Fig. 4c). The occurrence of the ($\pm 1/2$,0) intensity in the FT of the $dI/dV$ maps (Fig. 4f) is however subtler. By comparing the positions of the La2 shown as green dots (Fig. 4e) with this pattern in real space, we find a strong correlation, almost a one-to-one correspondence between the positions of La2 and the occurrence of the `stripy' pattern in the density of states maps. This indicates that the La2 atoms trap the same local 2a period resonance states as the La1 atoms. Unlike previous work where no connection was established between the La induced resonance states (6) and the glassy charge order, our data indicate that the La1 and La2 induced localized resonance states are responsible for much of the period 2a signatures in real and Fourier space. The connection between such local impurity resonance states and an incipient charge density wave instability remains an open interesting question. However, this signal is suppressed in the metallic sample and is therefore unlikely to be related to the DW instability (14) observed in this compound, which persists far into the metallic phase.

Interestingly, we find evidence for another electronic order in this system. Examination of the topographies at different bias voltages reveals that they are strongly energy dependent. At high negative bias voltages, the expected square lattice for Sr is observed (Fig. 1e). This suggests the Sr lattice remains undistorted despite the tilting of the oxygen octahedra. However, at other energies, a zigzag pattern is observed (Fig. 1f). Because topographic images contain information on both the surface corrugation as well as LDOS, the energy dependent topography we see here probably reflects charge rearrangement due an underlying electronic order. To visualize this electronic order, we utilize a supercell averaging algorithm (see Supplemental Material), which has been successfully applied to cuprates (25). The supercell algorithm is an averaging technique that utilizes the spatially dependent topography to construct an averaged unit cell that reflects any long-range ordered structure. From a series of bias dependent topographies on multiple samples, we plot the locations where maximal integrated DOS appears within unit cell as shown in Fig. 5d and also in the Supplemental Material.

**Discussion**

The binned-averaged spectra on samples with different doping allow us to make a few general observations on the effect of electron doping on the electronic structure. First as we go from x=0.032 to x=0.049 (Fig.3a-c) there is a clear overall trend in the spectral shapes. In the 0.032 and 0.039 samples there are still areas with insulating gaps although the largest gap magnitude observed has decreased significantly from ~130meV in the parent compound to 30meV in the 0.039 compound. Second, in the metallic samples, all spectra are gapless and show a similar overall spectral shape. We now concentrate on the 0.032 sample, and start with the position dependent changes of the conduction band edge near $E_F$. As shown in Fig. 3d, in areas with larger local densities of La2, the midpoint of conduction band leading edge moves towards the Fermi energy. However, the spectral shape of the conduction band does not change significantly (see inset to Fig. 3d where we have simply shifted the spectra in energy such that they lie on top of each other), suggesting the conduction band motion mimics a local potential variation or a local rigid band shift. This is consistent with photoemission data which show that with increasing La, the Fermi energy moves into the conduction band at the M point (17, 26). In a simple picture, the spatially varying leading-edge positions seen in STM data may be construed as an inhomogeneous distribution of charge causing potential variations about the mean doping level.

However, this is not the only effect of La2. As the local density of La increases, there is increasing spectral weight accumulation around -150meV (Fig. 3d), close to the hopping energy $t$~ 0.2eV (27). Simultaneously, the density of states at deeper energies (below -300meV) is suppressed. Thus, the sub-surface La is associated with a spectral weight transfer from the valance band towards the Fermi energy. This kind of spectral weight transfer is different from electron doping caused by missing anions on the surface (15, 28), where only the upper Hubbard band is involved, but is consistent with the theoretical models proposed for electron doped cuprates (29). It has been suggested that in the antiferromagnetic (AFM) insulating background, upon electron-doping, the creation of particle-hole spin excitations in the upper band strongly admixes the quasi-hole states in the upper and lower bands, giving rise to the emergence of spectral weight at an energy of approximately -$t$ in the charge gap. In $(Sr_{1-x}La_x)_3Ir_2O_7$, despite the destruction of static AFM, magnetic excitations can persist across the IMT (16, 18). Whether or not this treatment captures the nature of $(Sr_{1-x}La_x)_3Ir_2O_7$, the spectral weight transfer from the valence bands is a distinct aspect of electron doping in this system.

We now address possible reasons for dramatic differences between the effects of La1 and La2. In Ir214 (6), it was found that for low dopings below the IMT, the La atoms have a minimal effect on the local density of states due to Mott localization. For Ir214 this picture breaks down for dopings above 4%. (Note that Ir214 is a single layer compound and therefore has only one site for La dopants). One could extrapolate from that study and speculate that the different behavior of La1 and La2 is due to their different densities. In this scenario, the top layer with only 1.5% La behaves like the Ir214 sample below the IMT where the electrons are localized due to strong correlations. However, in the case of the bilayer, electrons from La2 are expected to reside in the $IrO_2$ planes, where they would contribute to screening the top SrO plane, thereby reducing the localization effect. The source of the difference between La1 and La2 is therefore not completely obvious and further experimental and theoretical work is necessary to pinpoint the origin of this dichotomy.

Finally, we discuss origins of the energy dependent topographies of the La-doped samples before metallicity sets in. While topographies at high negative energies below $E_F$ (crosses in Fig. 5d) reveal the Sr atoms at the expected positions of the lattice, the data in a band of low energies (below and above $E_F$) show a consistent shift of the LDOS with respect to Sr sites in alternate directions along the diagonal direction (Ir-Ir direction). This pattern, visually seen as a zig-zag pattern (fig. S7) represents the property of an underlying electronic order, which breaks the local two-dimensional inversion symmetry in the $IrO_2$ plane. Interestingly, the structural distortion, stemming from the tilting of $IrO_6$ octahedra (19) has the same symmetry as the electronic order we find here (Fig. 5e). To confirm that this order is tied to the tilting, we perform a comparative supercell analysis on the Ru-doped (B-site dopant) and pristine compounds where neither the DW nor the tilting has been observed. Correspondingly, the supercells in these compounds show no energy dependence (Fig. 5b and c; supplemental Fig.S8), and all the visible atoms are located at perfect square lattice sites as expected from the Sr lattice. In combination with the optical spectroscopy results, our data across the pristine, Ru-, and La- doped compounds indicate that this local inversion symmetry breaking charge order is intertwined with both the structural distortion and the DW.

**Conclusion**
Our measurements unveil the microscopic mechanism for the formation of a density wave state within the $J_{eff}=1/2$ state in $(Sr_{1-x}La_x)_3Ir_2O_7$. La dopants preferentially occupy the Sr sites in the middle plane of iridium oxide bilayers, dramatically changing the local electronic structure and inducing a spectral weight transfer from the lower Hubbard band to a new low-energy quasiparticle band. This leads to local electron density fluctuations that promote nanoscale phase separation prior to the formation of a global metallic state. The extended Coulomb interactions between electrons in the spatially large Ir 5d orbitals could play an important role (10). Within the new low energy band, multiple electronic orders emerge. A unidirectional order, which breaks local inversion symmetry, is absent in both the pristine, and B-site substituted samples, demonstrating that the charge instability is endemic to electron doping in this system. This primary charge order parameter intertwines with the previously reported structural distortion known to onset below 200K. We additionally observe local resonant states forming a checkerboard-like pattern trapped by La, which were also previously observed in $Sr_2IrO_4$ (6). Our combined data show a microscopic picture of how the weak Mott ground state in a bilayer iridate collapses with electron doping and indicates a potential coexistence of two types of order within one $J_{eff}=1/2$ band.

## Materials and Methods

**STM measurement** Single crystals of $(Sr_{1-x}La_x)_3Ir_2O_7$ were grown via flux techniques reported earlier (30), and the actual La concentrations were confirmed by energy dispersive spectroscopy (EDS). The IMT happens at x~0.04 and a detailed phase diagram can be found elsewhere (16). All samples are cleaved at ~77 K in ultra-high vacuum and immediately inserted into the STM head where they are held at ~4.3 K during the process of data acquisition. Differential conductance maps dI/dV(r, eV) and spectra dI/dV(eV), which are proportional to the LDOS at position *r* and energy eV, were obtained using standard lock-in technique.

**Average supercell algorithm** First, we use the drift-correction method (22) to remove the slow thermal drift and warping from piezo relaxation in the topographic images. The atoms are placed in the true position of the lattice after this treatment. Then we create a blank 2 × 2 supercell with better spatial resolution than the raw data. In this work, it is 25 × 25 pixels per unit cell. For each pixel in the drift-corrected topographies, we calculate its position in lattice coordinate, and then place it in the appropriate bin of the supercell. A detailed description of this method can be found in ref. 25. After 2 × 2 supercells obtained, we tile them two times to create a larger supercell for better visualization, as shown in Fig. 5.

## Acknowledgements

We thank Ilija Zeljkovic, Hsin Lin, Eduardo Fradkin and Milan Allan for helpful discussions. V.M gratefully acknowledges NSF Award No. DMR-1610143 for the STM studies. S.D.W. and T.H. acknowledge funding support from NSF Award No. DMR- 1505549.

## Competing financial interests

The authors declare no competing financial interest.

## Data Availability

The data sets generated during and/or analyzed during the current study are available from the corresponding author on reasonable request

**Figure 1**

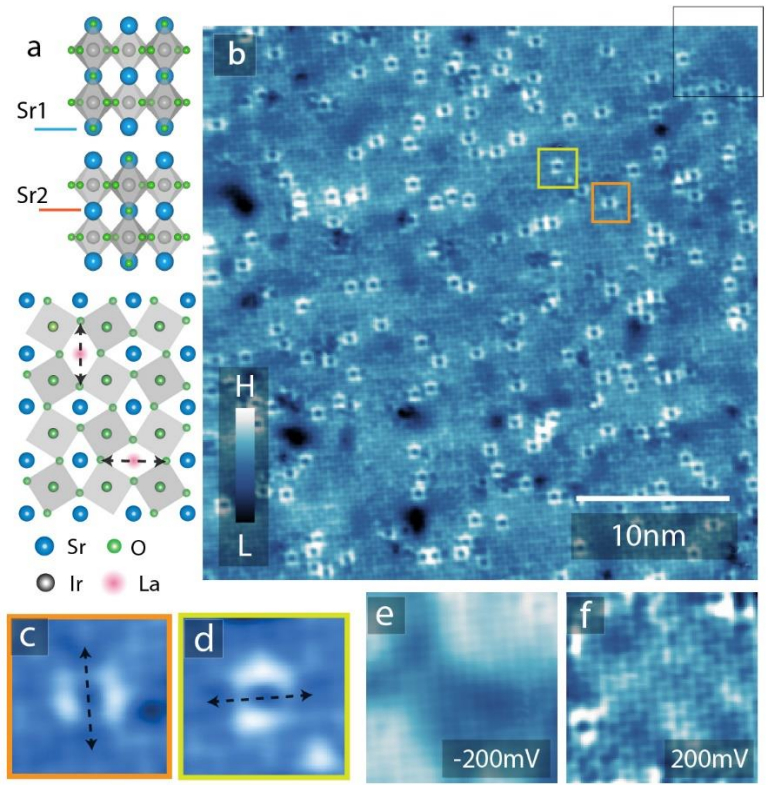

**Figure 1 Surface La dopants. a**. Crystal structure and schematic of the IrO$_6$ octahedral rotation in Sr$_3$Ir$_2$O$_7$. **b**. Atomic-resolution topography of La-doped sample with x=0.032. V$_S$=200mV, I$_t$=100pA. **c-d**. Zoom in on two types of La defects in **b** which show the reduction of local C4 symmetry to C2. **e-f.** Topographies taken with different biases in a 6.5nm*6.5nm area (marked with black square in **b**).

**Figure 2**

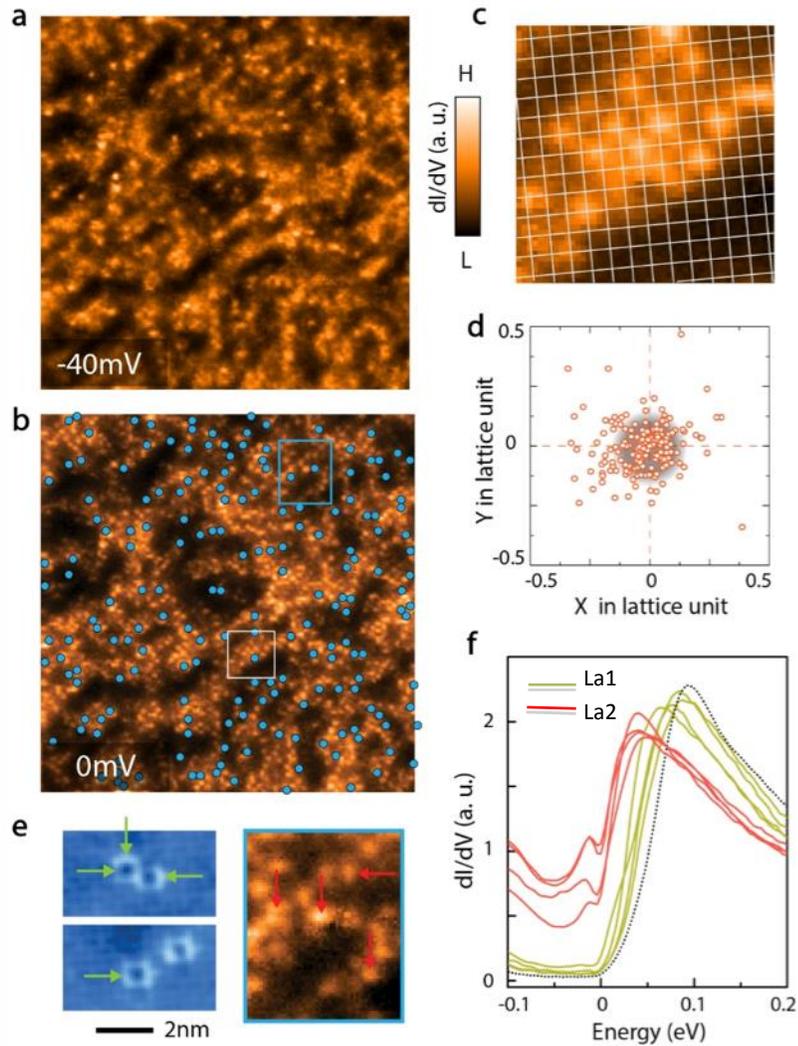

**Figure 2. Middle-layer La dopants. a-b**, dIdV maps taken in the same region of Fig. 1a at E= -40 and 0meV on the La-doped sample with x=0.032. Clear atomic scale features appear at the Fermi level (E= - 0meV). The blue dots denote the surface La1 dopants. **c.** Zoom-in of area marked by white square in **b**. The white lines denote the Sr lattice obtained from the simultaneously acquired topography. **d**. A scatter plot showing the positions of the bright features seen in the E= - 0meV dI/dV map, in lattice coordinates. The Sr position is set to (0,0). **e.** Images of La1 and La2. Arrows point to positions where the tunneling spectra shown in **f** were obtained. **f.** Tunneling spectra on La1 (green) and La2 (pink).

**Figure 3**

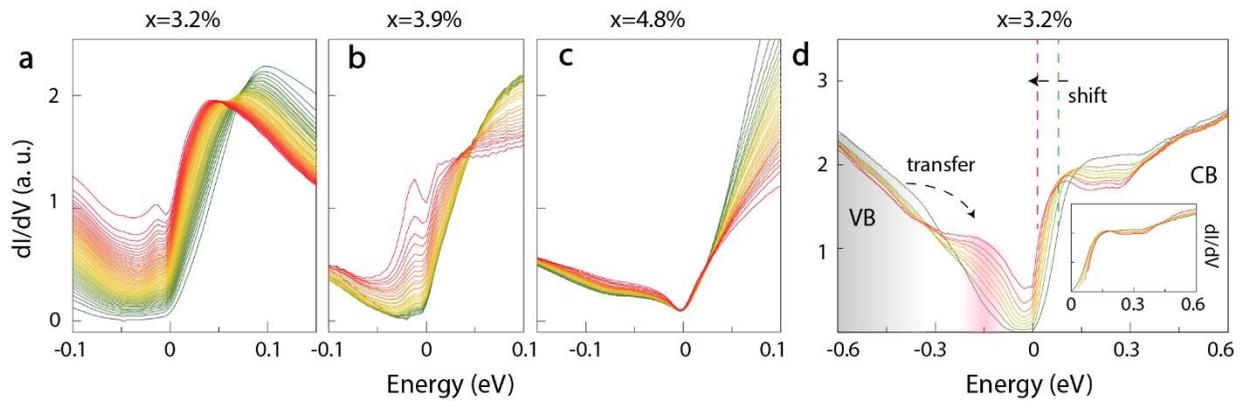

**Figure 3. Evolution of tunneling spectra across IMT. a-c**. Binned spectra for x=0.032, 0.039 and 0.048 samples. Differential tunneling conductance at -40, -40, and 100meV were chosen to bin the spectra for a, b and c, respectively. Green/red curves denote more insulating/metallic behavior in **a** and **b**. **d**. binned spectra in a larger energy range on the x=0.032 samples. Inset: normalized spectra with shifted energy to show that the overall spectral shapes above Fermi level overlap well after shifting in energy.

**Figure 4**

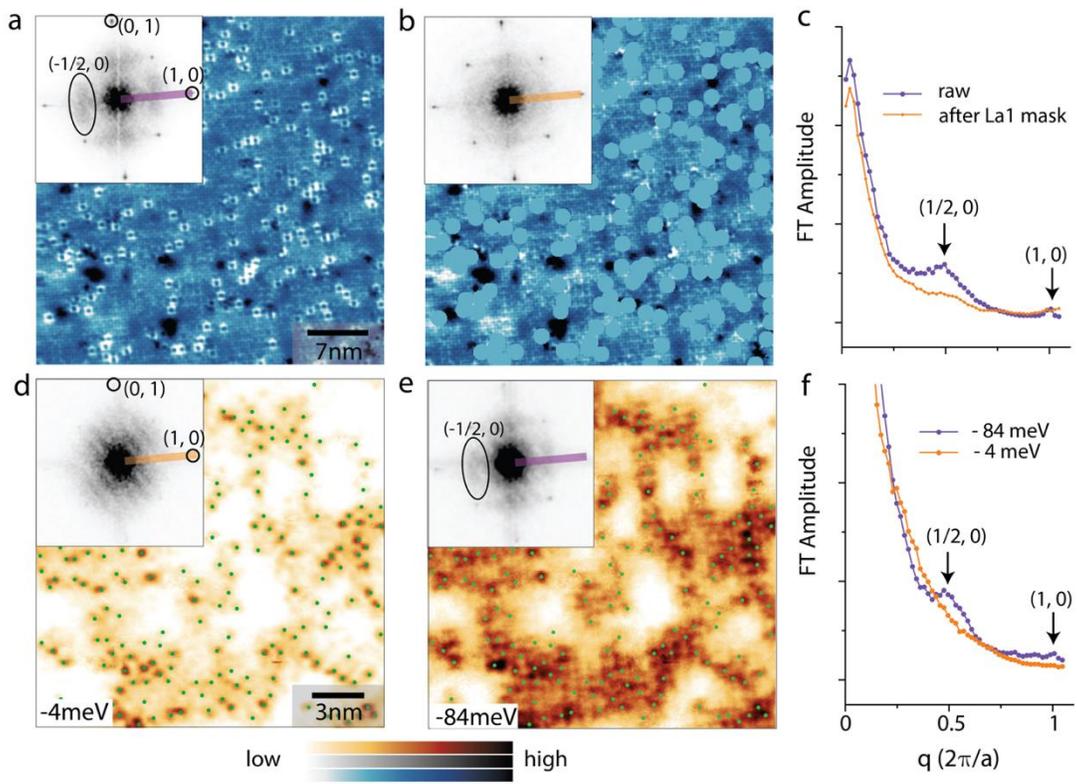

Figure 4. Features in FT at (1/2, 0). **a.** Atomic-resolution topography of La-doped sample with x=0.032 (same as Fig.1b). The inset shows its FT. (0,1) and (1,0) are the Sr reciprocal lattice vectors. **b.** Masked topography to remove the La1 features in a and its FT. **c.** Linecuts along (1,0) direction of the FT before and after masking La1. **d, e.** dI/dV map at -4 meV and -84meV, respectively. The green dots denote the position of La2. dI/dV map at -84meV shows glassy checkerboard-like modulation, clearly centered around the locations of La2. **f.** Linecuts along (1,0) direction taken on the FTs shown in the insets in d and e.

# Figure 5

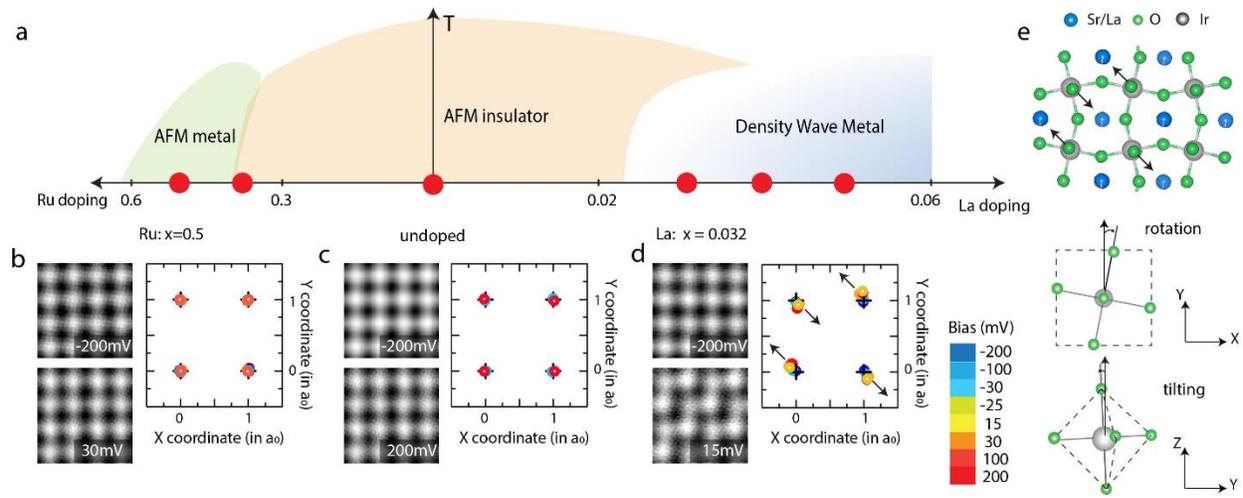

**Figure 5. Visualizing the density wave a.** Schematic phase diagram for La and Ru doping in bilayer Ir327. Red dots represent the doping levels we studied here. **b-c**, Energy dependent averaged supercell in 4 ×4 blocks for pristine, Ru-doped and La doped samples. The center of visible 'atom' is fit by 2D Gaussian function and plotted in unit cell coordinate. The charge distribution shifts away from the Sr lattice at low energies for La doped samples. **e,** Schematic of the IrO$_6$ octahedra tilting in La doped samples (x=0.035). Lower panel: cartoon images showing in-plane rotation and off-plane tilting of the IrO$_6$.

# Supplementary information for
# "Doping induced Mott collapse and density wave instabilities in $(Sr_{1-x}La_x)_3Ir_2O_7$"

Zhenyu Wang, Daniel Walkup, Yulia Maximenko, Wenwen Zhou, Tom Hogan, Ziqiang Wang, Stephen D. Wilson and Vidya Madhavan[*]

**This PDF file includes:**

**La1 dopants in Sr327**

**La doping concentration**

**La2 images at different energies**

**Spectroscopic linecuts across La1 and La2**

**Correlation between the electronic inhomogeneity and La1 and La2 dopants**

**Global metallic state in x=4.8% samples**

**Average supercell algorithm**

**Fig.S1: Surface La1 dopants on Sr sites.**

**Figure S2. La doping concentration**

**Figure S3. La2 dopants in dI/dV maps.**

**Figure S4. Linecuts across La1 and La2 dopants.**

**Figure S5. Electronic inhomogeneity and La dopants.**

**Figure S6. Line-cut in x=4.8% samples.**

**Figure S7. Average supercell: La-doped Sr327.**

**Figure S8. Average supercell: undoped and Ru-doped Sr327.**

# La1 dopants in Sr327

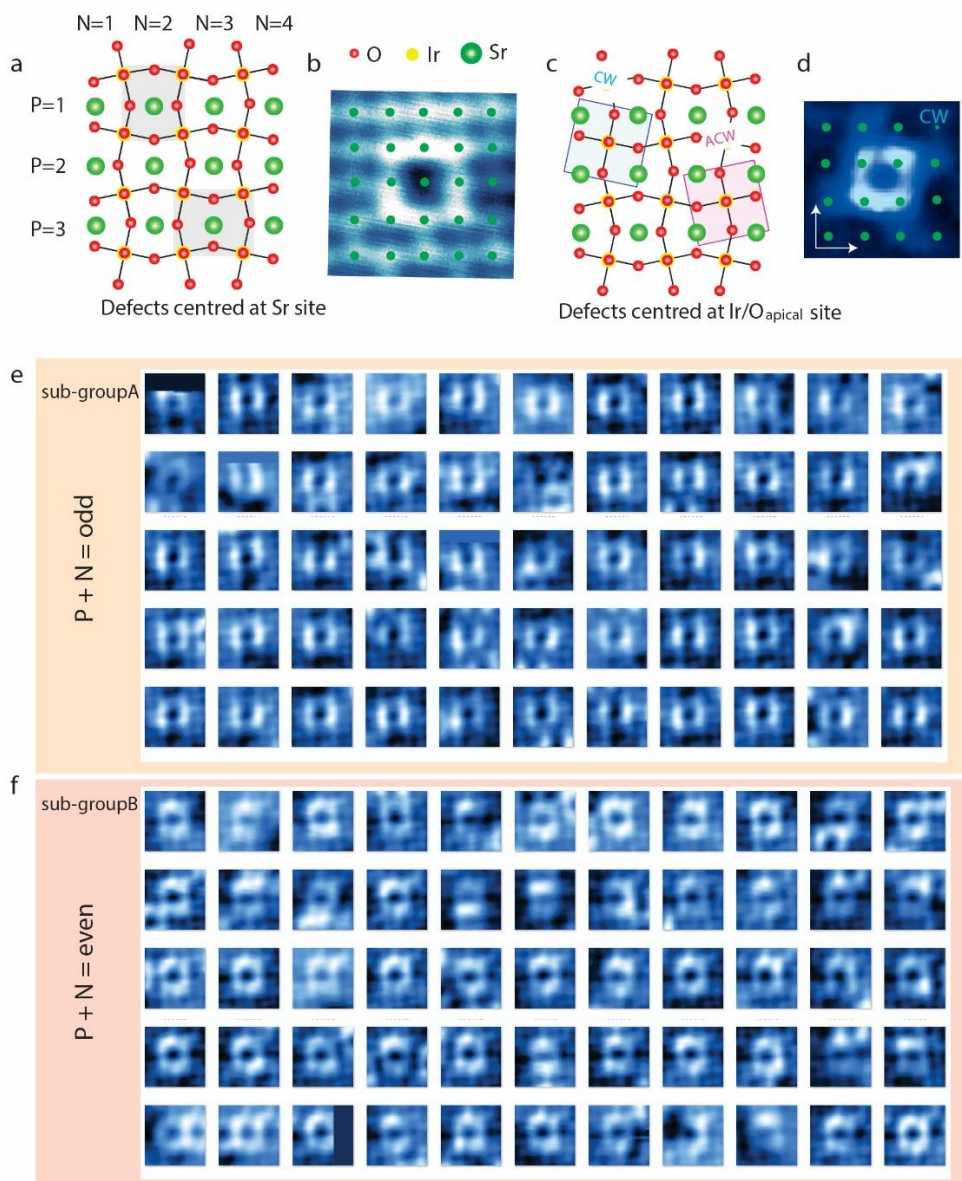

**Figure S1. Surface La1 dopants on Sr sites.** a, c) Schematic top views showing the local symmetry of Sr sites (a) and Ir/ apical oxygen sites (c), respectively. Figures are reproduced from [1]. b, d) Surface La dopant (La1) centred on Sr site and apical oxygen vacancy from our earlier study of the parent compound centred in between four Sr sites. The green dots denote Sr sites. e, f) Two sub-groups of La dopants: all the 'left-right' defects are on odd-numbered Sr sites while all the 'up-down' ones are on even-numbered sites.

The surface atomic spacing in our topographies is about 0.39nm, which can be explained equally well by Sr-Sr or apical O-O distance. Usually it is difficult to tell which lattice we see on the surface, however in Sr327, the staggered rotation of $IrO_6$ octahedra provides an obvious clue: as shown in Fig.S1a and c, the environment of a Sr site preserves refection symmetry but breaks C4 rotation symmetry, while for apical oxygen/Ir site it is exact opposite. In our images, the surface La dopants (La1, Fig.S1b), centred on the topographic maxima, are clearly dimorphic, while the chiral apical oxygen vacancies which break refection symmetry are centered in between four maxima (Fig.S1d). These observations allow us to confirm that the lattice in topography corresponds to Sr. By labelling the Sr sites, we find that two type of La1 dopants are perfectly correlated with the lattice positions. Some of the results are shown in Fig. S1e and f.

# La doping concentration

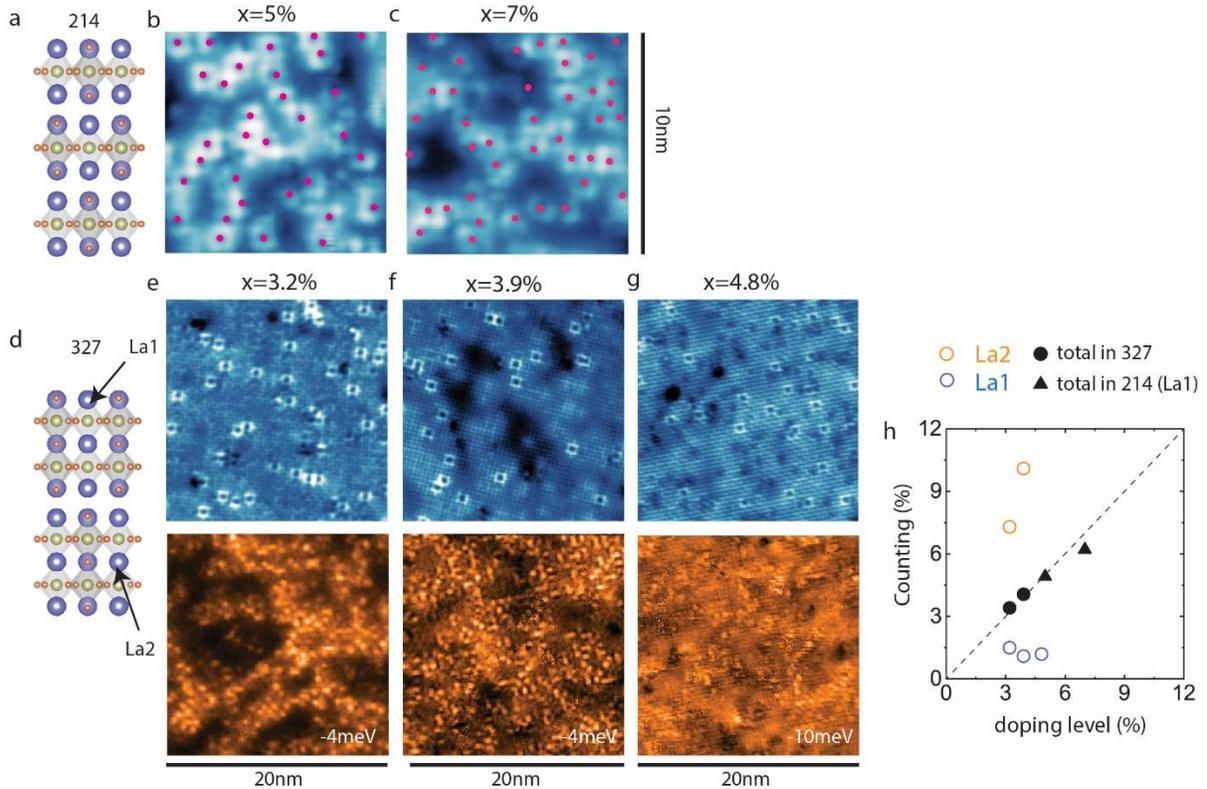

**Figure S2. La doping concentration.** a, d) Crystal structures of Sr214 and Sr327. b, c) Topographies of 5% and 7% La-doped $Sr_2IrO_4$. The pink dots denote the positions of La dopants on the surface. e-g) Topographies (upper panels) and dI/dV maps near Fermi level (lower panels) for La-doped $Sr_3Ir_2O_7$. (h) Summary of the La concentrations in different samples.

In Fig. S2a-c, we show two topographies obtained on 5% and 7% La-doped Sr214 samples. In contrast to Sr327, there is no middle SrO layer in the unit cell of Sr214, so the number of observed La dopants on the surface does correspond to the nominal doping level of the samples. In Sr327, La dopants can hide in the middle SrO layer so the total value should include both La1 and La2 sites (Fig. S2d-g). For example, in x=3.9% samples, we count 28 La1 (1.1%) and about 270 La2 (10.1%) in a 20nm*20nm area, so the total La concentration is about 1/3*10.1%+2/3*1.1%= 4.1%. As a summary, we show the La concentration of different samples in Fig. S2h, where one can see the counting concentrations of La are consistent with the actual doping level both in Sr214 and Sr327.

# La2 images at different energies

To have a better understanding of La2 dopants, we present energy dependent dI/dV maps in Fig. S3. Near Fermi level, La2 dopants appear as bright dots in the differential conductance map. However, at some certain energies, the signature of La2 dopant becomes square-shaped, which is very similar to that of a surface La1 dopant in topography. This observation suggests that La2 dopants are more likely to be substitutions other than interstitial atoms because they share a similar potential symmetry with La1.

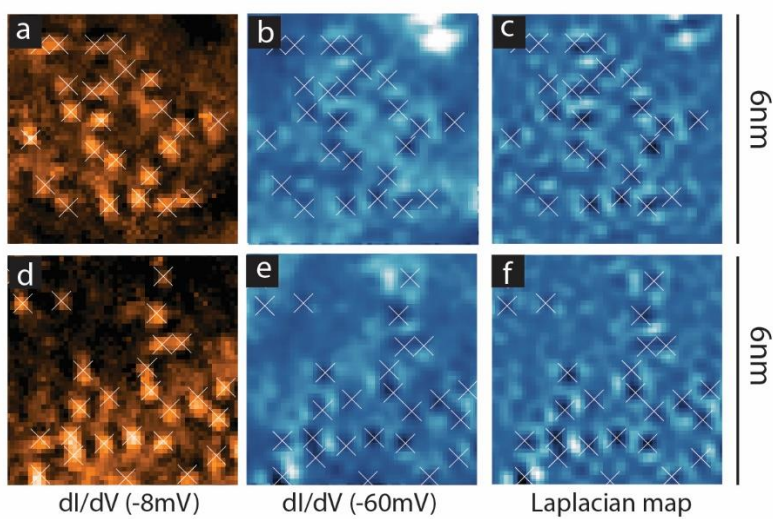

**Figure S3. La2 dopants in dI/dV maps.** a, d) dI/dV maps in two 6nm*6nm ranges at -8meV. b, e) dI/dV maps in the same range as (a) and (d) at -60meV. One can see the bright dot has a square-like shape which is similar to the shape of La1 in the topography. c, f) Negative Laplacian maps of (b) and (e) to highlight the square-shaped features.

# Linecuts across La1 and La2

Figure S4c and f show two line-cut profiles of dI/dV spectra across La1 and La2 respectively. The positions of La1 and La2 dopants have been carefully selected (far away from each other) to isolate their effects on the local density of states. The effect of La2 (bright spots) is much more dramatic than La1 (surface square).

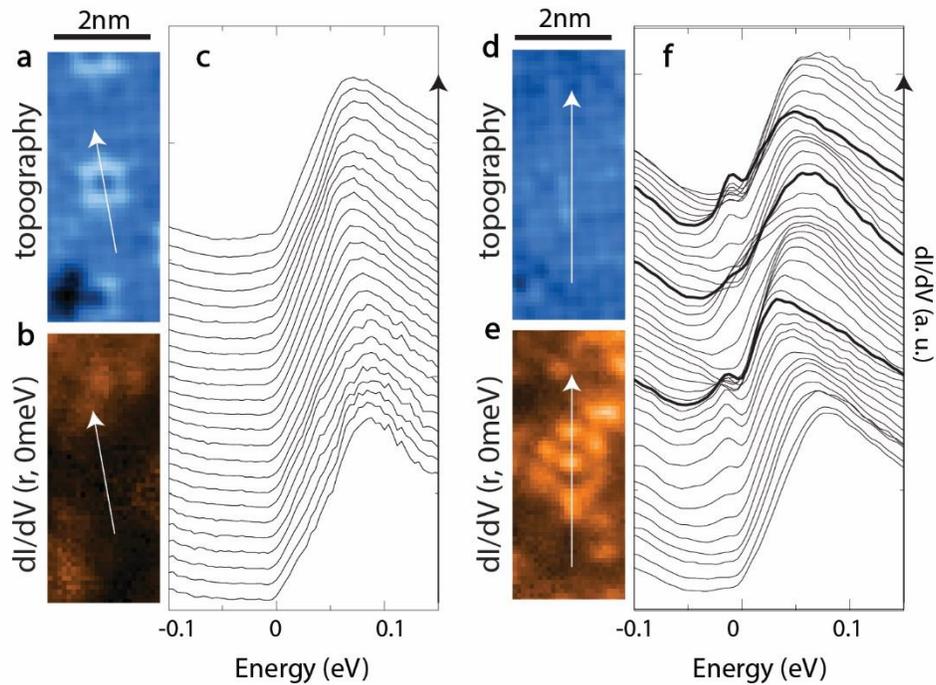

**Figure S4. Linecuts across La1 and La2 dopants.** Left: line cut across La1 (square on the surface). Right: line cut across three La2 dopants (bright spots in dI/dV maps).

# Correlation between the electronic inhomogeneity and La1 and La2 dopants

To establish a connection between the electronic inhomogeneity and different type of La dopants, we perform a cross-correlation analysis in a 20nm* 20nm area. First, we use dI/dV (r, 0meV) (Fig. S5a) and topography (Fig. S5d) to address the position of La2 dopants (Fig. S5b) and La1 dopants (Fig. S5e), respectively. Then the effects of these dopants are considered as 2D Gaussian function centered at each dopant. For example, Fig.S5c and f show the images obtained with a full width half maximum (FWHM) of about 0.52nm for La2 and La1, respectively. Finally, we do a cross-correlation between dI/dV (r, -40meV) (Fig. S5g) and dopants distribution images with different FWHM. The result is shown as yellow and blue dots in Fig. S4h, which also suggests that the electronic inhomogeneity is mostly associated with the presence of La2. We also see very weak correlation between La1 and La2 (pink dots in Fig. S4h).

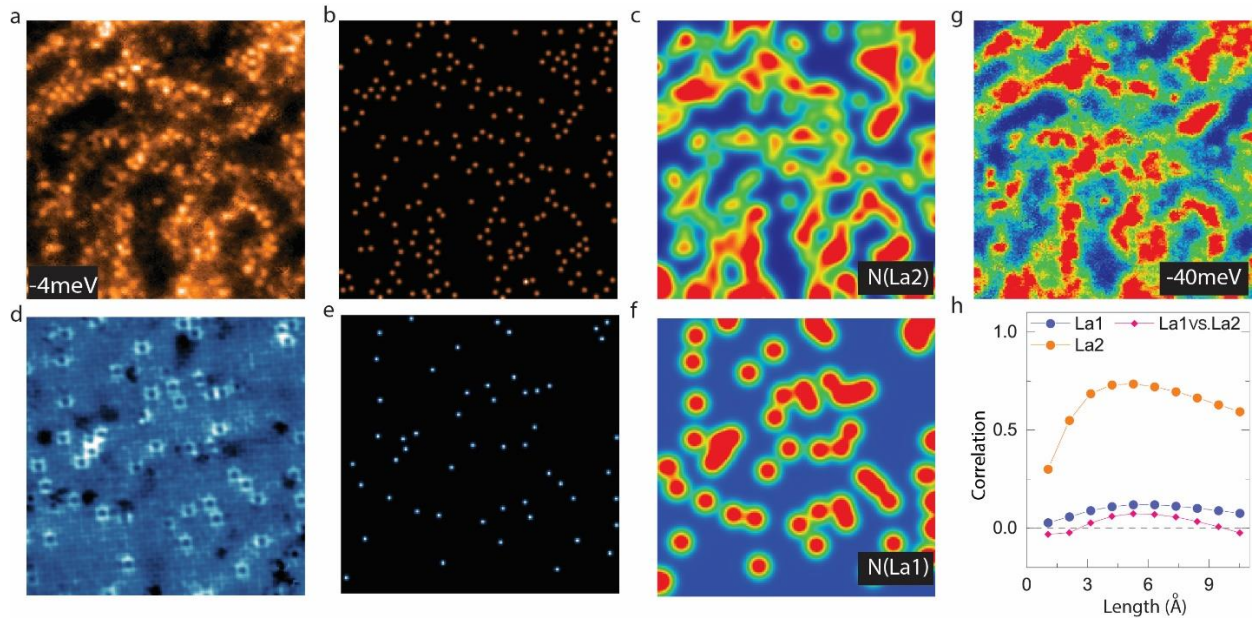

**Figure S5. Electronic inhomogeneity and La dopants.** a) dI/dV (r, -4meV) where La2 can be observed clearly. b) the position of La2 in (a) are marked with bright spots. c) La2 distribution images obtained with FWHM about 0.52nm. d-f) similar analysis for La1 dopants from topography. g) dI/dV (r, -40meV) showing clear insulating-metallic phase separation. h) correlation between (g) and the dopants distribution images with various FWHM.

# Global metallic state in x=4.8% samples

In x=4.8% compounds, the spectra are gapless everywhere and shows a uniform 'V' shape (Fig. S6 and Fig. 3c), which is consistent with the phase diagram given in [2]. This 'V' shape is a little asymmetric with respect to Fermi level, where the positive side is steeper. In many of the spectra, there is also a small feature at around -16meV.

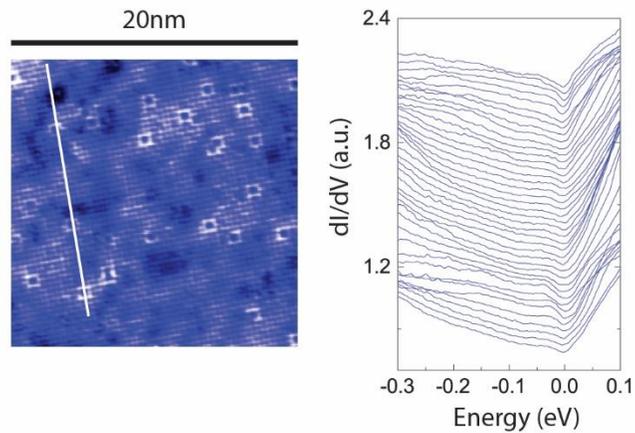

**Figure S6. Line-cut in x=4.8% samples.** Left: topography showing where the line cut was obtained. Right: Tunneling spectra along the line.

# Average supercell algorithm

We first present a brief introduction of the average supercell algorithm used in this work. More details can be found in [4][5]. First, we use the drift-correction method [6] to remove the slow thermal drift and piezo relaxation in the topographic images. We should note that we only use drift-correction to correct large length scale distortions that occur over many lattice constants. Then, for each pixel in this drift-corrected image, we calculate its position in lattice coordinate, and place it in the appropriate bin in the 2*2 supercell. The average supercell is a histogram with much higher spatial resolution (usually can improve the resolution from 3-4 pixels per unit cell to 25 pixels per unit cell). After the 2*2 supercells are obtained, we tile them two times to create a larger 4*4 supercell for better visualization (Fig.5 in main text, Fig. S7 and S8).

We apply the average supercell algorithm to our STM topographies of different samples in Sr327 family. Fig. S7 and S8 show series of average supercells as a function of scanning bias in pristine, Ru-doped and La-doped Sr327 samples. For the former two cases, the average unit cells show bright maxima whose positions do not change with bias. This is confirmed in the coordinate plots, which represent the centers of these averaged maxima as determined by an2D Gaussian fit. As shown in the coordinate plots, in the pristine and Ru-doped compounds (Fig. S8), all the bright maxima perfectly align on the Sr lattice as expected over a large energy range.

The La doped samples are different and show a striking energy dependence in the topography. Let's study the 3.2% sample first as shown in Fig. S7a. At high negative scanning bias (-Vs > 50mV), the supercells look like the expected Sr square lattice and indeed, the coordinate positions match the expected Sr lattice. However, as we approach the Fermi level, we observe a displacement of the bright maxima away from the Sr sites (clearly seen at -25mV or 15mV for example), which persists into the positive side.

The STM topographic image is a convolution of the surface corrugation and electronic density of states, therefore the bright maxima seen in the average unit cell do not only represent the atomic positions. Since the positions of the Sr atoms are not expected to move with the bias, the distortion we see starting at low negative bias and going across the Fermi energy, must be a result of charge rearrangement. This suggests that there exists a density-wave-like electronic instability near Fermi level, which moves the maxima away from Sr site in our topographic images. This kind of distortion can be observed in all the La-doped samples studied here (x=3.2%, 3.9% and 4.8%). Note that a slightly elongated tip shape can affect the perceived order as shown in the cartoon in Fig. S7 e and f. This explains why the charge order at 4.8% doping looks different to the eye (the 3.9% at 300mV looks like **f** while the other two dopings look like **e**). However, the distortion is still the same in the 4.8 % sample as can be seen in the coordinate plots.

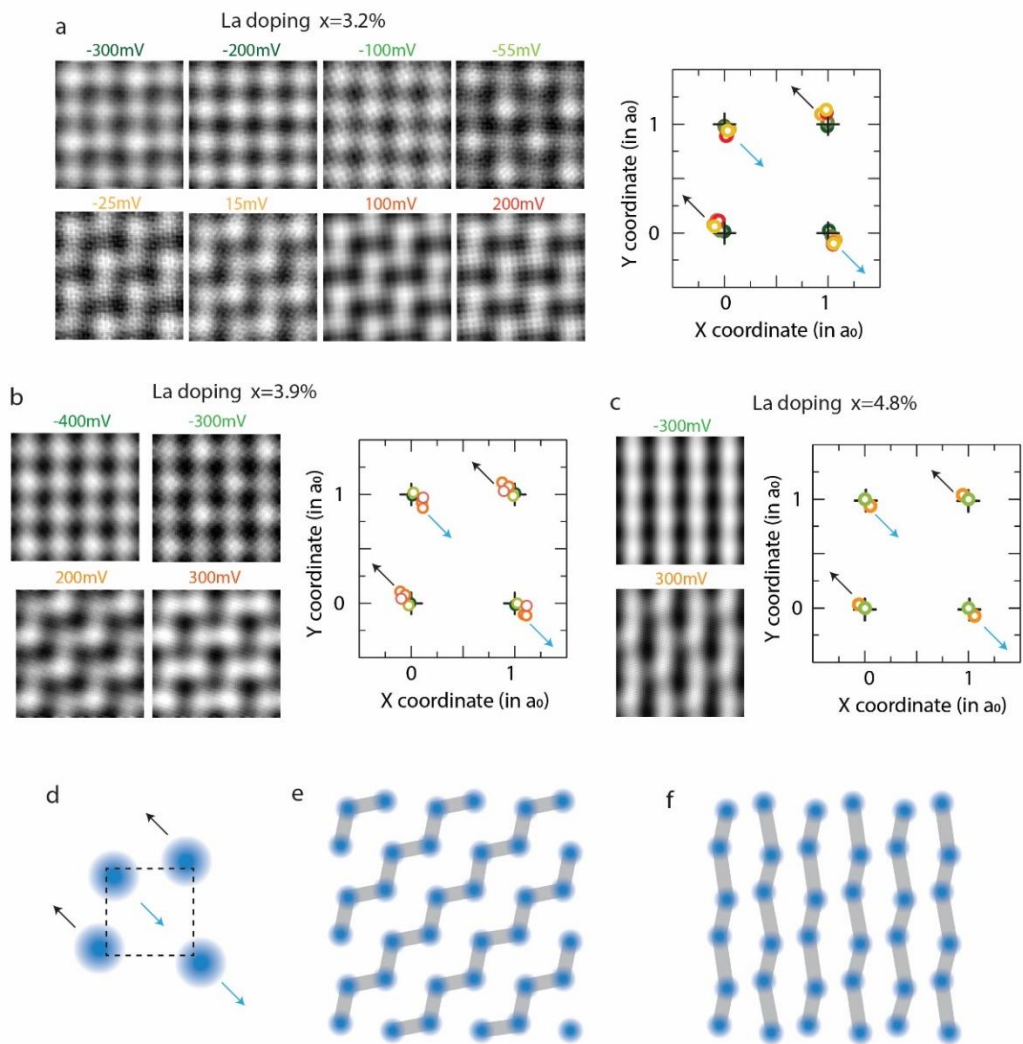

**Figure S7. Average supercell: energy dependence of topographic maxima in La-doped Sr327.** Averaged supercell obtained in a 44nm*44nm area of the x=3.2% La doped Sr327 (a); in ~20nm*20nm areas of the x=3.9% La-Sr327 (b); in a 30nm*30nm area of the x=4.8% La- Sr327 (c). d)-f) Schematic of the charge order. (e) and (f) show *identical* charge order as represented by the blue dots. However, while in most cases the charge order looks like (e), a slightly elongated tip, as we have for the 4.8% sample, can make the charge order look like (f).

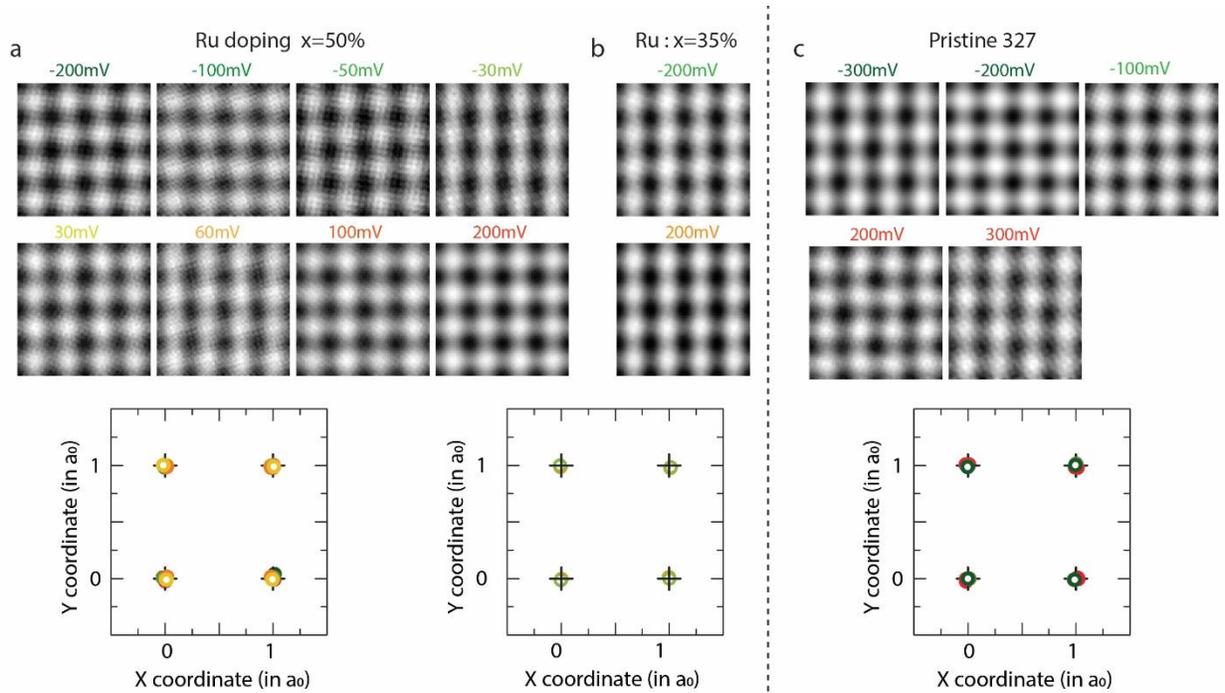

**Figure S8. Average supercell: energy dependence of topographic maxima in undoped and Ru-doped Sr327.** Averaged supercell obtained in about 20nm*20nm areas of the x=50% Ru doped Sr327 (a), in a 30nm*30nm area of the x=35% Ru doped Sr327 (b) and in a 30nm*30nm area of Pristine Sr327 (c).

## Supplementary References